\documentclass{jpsj3}
\usepackage{txfonts}

\title{Superconductivity in La$_3$Pt$_4$}

\author{\name{Yuki \surname{Kawashima}}\thanks{E-mail address: kawashima@scphys.kyoto-u.ac.jp}, \name{Gaku \surname{Eguchi}}, \name{Shingo \surname{Yonezawa}}, and \name{Yoshiteru \surname{Maeno}}
}
\inst{\address{Department of Physics, Graduate School of Science, \\ Kyoto University, Kyoto 606-8502} 
}

\kword{superconductivity, La$_3$Pt$_4$}

\begin{document}
\maketitle


We found superconductivity of intermetallic La$_3$Pt$_4$ below 0.51~K, 
while searching for the Pt-substituted material of the noncentrosymmetric (NCS) superconductor LaNiC$_2$ 
($T_{\rm{c}}=2.7$~K\cite{Lee_96, Pecharsky_98}). 

\begin{figure}
\begin{center}
\includegraphics[width=9cm]{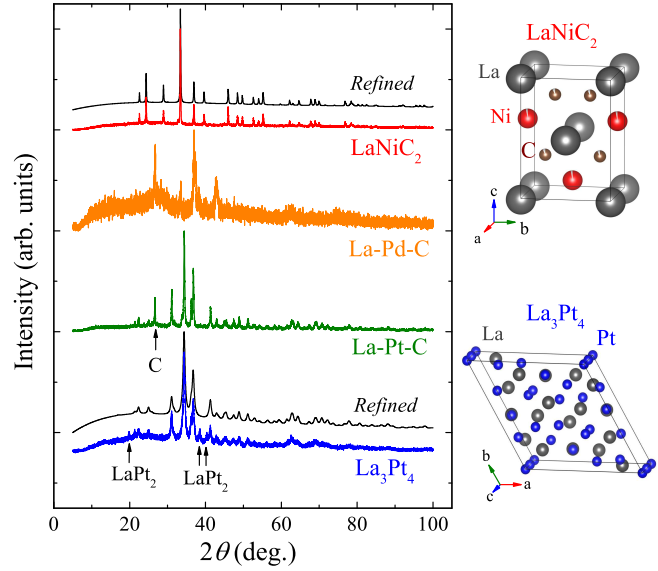}
\end{center}
\caption{
(Color online) Powder XRD patterns of LaNiC$_2$, La-Pd-C, La-Pt-C, and La$_3$Pt$_4$ samples 
(Cu K$\mathrm{\alpha}$ radiation). 
The two patterns in black show the refined XRD patterns obtained by the Rietveld method. 
The impurity peaks of carbon (in La-Pt-C) and LaPt$_2$ (in the La$_3$Pt$_4$ sample) are indicated. 
The right figures show the crystal structures of LaNiC$_2$ and La$_3$Pt$_4$. 
These schematics are drawn using the software VESTA\cite{VESTA_11}. 
}
\label{fig_XRD}
\end{figure}

Superconductors with NCS crystal structures, which do not have spatial inversion symmetry, 
have been attracting much attention since the discovery of 
the absence of the Pauli limiting field in CePt$_3$Si\cite{Bauer_04}. 
Recently, some interesting superconducting properties were reported in LaNiC$_2$\cite{Lee_96}. 
This compound crystallizes in the CeNiC$_2$-type orthorhombic structure (space group $Amm$2), which lacks 
the inversion symmetry along the $c$-axis (see Fig. \ref{fig_XRD}). 
A recent muon spin relaxation ($\mu$SR) result suggests unconventional superconductivity\cite{Hillier_09}. 
On the other hand, first-principle calculations for this material 
suggested that it is likely to be a
conventional superconductor judging from its s-orbital-dominant 
electronic states near the Fermi energy\cite{Subedi_09}. 
Further confirmation of unconventional properties is needed for this material. 

In NCS superconductors, antisymmetric spin-orbit interaction (ASOI) due to the 
absence of spatial inversion symmetry may lead to a spin singlet-triplet mixed superconducting state. 
A number of unconevntional superconducting properties are predicted, but 
most of them are yet to be observed\cite{Fujimoto_07}. 
Since stronger ASOI is considered essential for the emergence of the unconventional superconducting 
properties, it is desirable to investigate 
NCS superconductivity in compounds containing heavy elements. 
For LaNiC$_2$, replacements of Ni with heavier elements such as Pd or Pt seem favorable to 
realize large ASOI. 

We tried to synthesize La$M$C$_2$ ($M$ = Ni, Pd, Pt) samples using 
the arc melting method. 
The starting materials were La (purity 99.9\%), Ni (99.999\%)
, C (99.999\%), Pd (99.95\%), and Pt (99.98\%). 
These materials were melted in argon with the stoichiometric ratio of 
La:$M$:C = 1:1:2. 
Powder X-Ray diffraction (XRD) measurements (Bruker AXS, D8 ADVANCE) 
were carried out (Fig.~\ref{fig_XRD}). 
As represented in Fig.~\ref{fig_XRD}, the samples for $M$ = Ni are almost single-phase LaNiC$_2$. 
The XRD patterns of La-Pd-C and La-Pt-C samples do not resemble 
that of LaNiC$_2$. 
Further XRD analysis revealed that carbon remains unreacted in La-Pt-C samples, 
and that its main constituent is actually La$_3$Pt$_4$, not ``LaPtC$_2$''. 
La$_3$Pt$_4$ crystallizes in the rhombohedral Pu$_3$Pd$_4$-type structure 
(space group R$\bar{3}$\cite{Palenzona_77}, see Fig.~\ref{fig_XRD}). 
Note that its crystal structure has inversion symmetry. 

New samples were synthesized by arc melting from stoichiometric mixture of 
La:Pt = 3:4 for La$_3$Pt$_4$. 
As presented in Fig.~\ref{fig_XRD}, the XRD pattern of a La$_3$Pt$_4$ sample well agrees with that of 
La-Pt-C. 
The XRD indicates that its main phase is certainly La$_3$Pt$_4$ 
although it contains a small amount ($<$20\%) of LaPt$_2$ ($T_{\rm{c}}$ = 0.46 K\cite{Matthias_65}) impurity. 
Except for LaPt$_2$, no impurity phases are detected in this sample.  
The lattice parameters of La$_3$Pt$_4$ obtained using the Rietveld analysis were 
$a = 13.8$~\AA, $c = 5.83$~\AA, 
which agree with the literature\cite{Palenzona_77}.

\begin{figure}[tbp]
\begin{center}
\includegraphics[width=9cm]{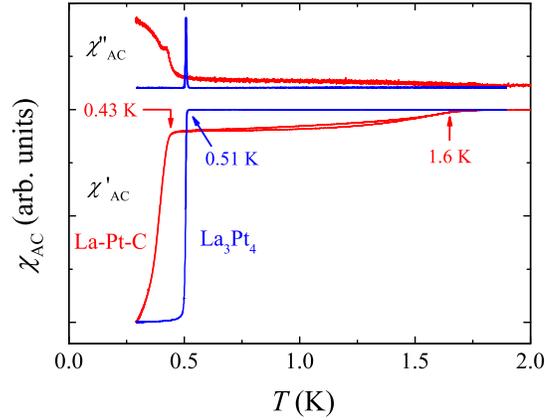}
\end{center}
\caption{
(Color online) Temperature dependence of the real and imaginary parts of the 
AC magnetic susceptibility $\chi_{\mathrm{AC}}$ 
of the samples of La-Pt-C and La$_3$Pt$_4$ down to 0.3 K. 
The AC magnetic field $H_{\mathrm{AC}}$ was 1 $\mathrm{\mu}$T-rms, and its frequency was 887 Hz 
for the La-Pt-C sample and 3011 Hz for the La$_3$Pt$_4$ sample. 
}
\label{fig_chiAC}
\end{figure}

The AC susceptibility $\chi_{\rm{AC}}$ of the synthesized La-Pt-C sample and La$_3$Pt$_4$ 
is presented in Fig.~\ref{fig_chiAC}. The measurements were performed by a mutual-inductance technique 
with a home-built first-derivative coil, mounted in a commercial $^3$He refrigerator 
(Oxford Instruments, Heliox). 
For La$_3$Pt$_4$, strong magnetic shielding indicating bulk superconductivity 
is observed below 0.51~K. 
Similar shielding signal is also observed in the La-Pt-C sample, indicating that La$_3$Pt$_4$ is 
indeed the majority phase in this sample. 
In addition, weak magnetic shielding below 1.6~K is observed only in the La-Pt-C sample. 
The shielding is presumably a superconducting impurity phase, 
but the origin is unclear. Note that LaC$_2$ is known to exhibit superconductivity 
below 1.6~K~\cite{Green_69}, although the corresponding XRD peaks are not detected.

\begin{figure}[tbp]
\begin{center}
\includegraphics[width=9cm]{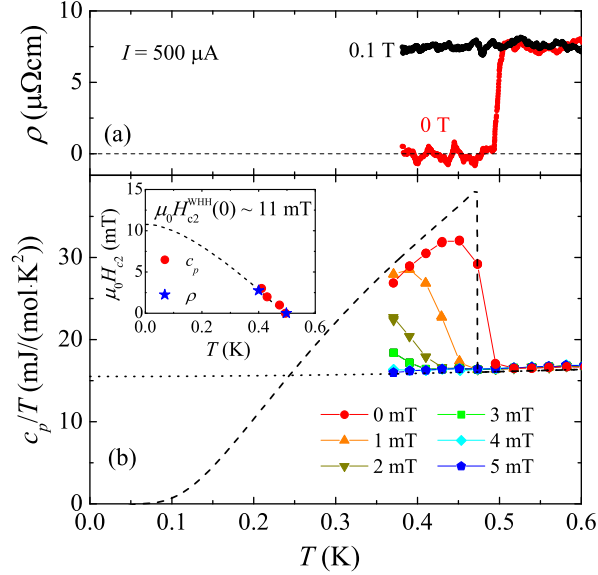}
\end{center}
\caption{
(Color online) 
(a) Temperature dependence of $\rho$ of La$_3$Pt$_4$ under 0~T and 0.1~T at low temperatures. 
(b) Temperature dependence of $c_p/T$ of La$_3$Pt$_4$ at low temperatures. 
The dotted line is a fit of $c_p/T = \gamma + \beta T^2$ to the normal state data up to 2.0 K. 
The obtained parameters are 
$\gamma = 15.5$~mJ/(f.u.mol$\cdot$K$^2$) and $\beta = 2.40$~mJ/(f.u.mol$\cdot$K$^4$). 
The curve based on the conventional BCS theory deduced from $\gamma$ and $T_{\rm{c}} = 0.47$ K 
is also presented\cite{1959_Muhlschlegel}. 
The inset shows the upper critical field $H_{c2}(T)$ 
determined from the onset $T_{\rm{c}}$ of $c_p/T$ (red circles) and 
the resistive midpoint (blue stars). 
The dashed line is the WHH curve for $H_{\rm{c2}}(T)$ in the dirty limit\cite{1966_WHH},  
giving $\mu_0H_{\rm{c2}}^{\rm{WHH}}(0)=
-0.693T_{\rm{c}}[{\rm{d}}(\mu_0H_{\rm{c2}})/{\rm{d}}T]|_{T=T_{\rm{c}}}=11$~mT. 
}
\label{fig_Cs}
\end{figure}

The resistivity of La$_3$Pt$_4$ under 0~T and 0.1~T is presented in Fig.~\ref{fig_Cs}(a). 
It was measured by a conventional DC four-probe technique with Heliox as well. 
The zero-resistivity temperature 0.5~K well agrees with the onset temperature 
of $\chi_{\rm{AC}}$. 
The residual resistivity ratio $RRR \equiv \rho_{300{\rm{K}}}/\rho_{1{\rm{K}}}$ is approximately 30 
(not shown) for the present polycrystalline samples. 
The specific heat $c_p$ divided by temperature in several magnetic fields is 
presented in Fig.~\ref{fig_Cs}(b). 
The heat capacity was measured using a relaxation-time-method 
with a commercial apparatus (Quantum Design, PPMS) down to 0.35~K. 
The molar specific heat are evaluated assuming single-phase La$_3$Pt$_4$. 
The clear and large specific heat jump at 0.5~K in 0~T due to the transition indicates 
the bulk superconductivity. 

Superconductivity is suppressed by a magnetic field of 0.1~T, as evidnced by both $\rho$ and $c_p/T$ 
measurements. 
The normal state $c_p/T$ is independent of the applied magnetic field within 
the experimental resolution. 
Thus $c_p/T$ at 0.1~T can be used to evaluate specific-heat coefficients: 
$c_p/T$ in the normal state is fitted by the relation $c_p/T = \gamma + \beta T^2$, 
where $\gamma$ and $\beta$ are the electronic and the phononic specific-heat 
coefficients, respectively. 
The value of $\gamma$ is evaluated to be 15.5~mJ/(f.u.mol$\cdot$K$^2$). 
We also calculated $\gamma$ based on the first-principle calculation using the 
WIEN2k package\cite{2003_WIEN2k} as 3.0~mJ/(f.u.mol$\cdot$K$^2$). 
Thus, the electronic mass enhancement is approximately 5. 
The obtained $\beta$ value is 2.4~mJ/(f.u.mol$\cdot$K$^4$), yielding the Debye temperature 
$\varTheta_{\rm{D}}=178$~K from the relation 
$\beta=(12/5)\pi^4N_{\rm{A}}N_{\rm{f.u.}}k_{\rm{B}}/\varTheta_{\rm{D}}^3$. 
Here $N_{\rm{A}}$ is the Avogadro number, $N_{\rm{f.u.}}=7$ is the number of atoms per formula unit, 
and $k_{\rm{B}}$ is the Boltzmann constant. 

The specific-heat jump height $\Delta c_p$ divided by $\gamma T_{\rm{c}}$ is smaller than 
that of the weak-coupling BCS theory: 1.43. 
The fact suggests existence of residual density of states that does not contribute to 
the superconductivity. 
The residual density of states is attributable to impurity phases that were detected 
in the X-ray spectrum in Fig.~\ref{fig_XRD}. 
Improvement of sample quality as well as measurements down to a lower-temperature 
region is necessary for further investigation. 

The superconducting $H-T$ phase diagram based on the onset $T_{\rm{c}}$ of $c_p/T$ and the 
resistive midpoint is presented in the inset of Fig.~\ref{fig_Cs}(b). 
The conventional Werthamer-Helfand-Hohenberg (WHH) curve\cite{1966_WHH} 
with the onset $T_{\rm{c}}=0.5$~K and the initial slope 
$[{\rm{d}}(\mu_0H_{\rm{c2}})/{\rm{d}}T]|_{T=T_{\rm{c}}}=32$~mT/K 
is also shown. 
The WHH upper critical field is estimated to be 11~mT. 
From this value, we crudely estimate the zero-temperature GL coherence length $\xi(0)$ 
as 170~nm, using the orbital depairing relation 
$\mu_0H_{\rm{c2}}^{\rm{WHH}}(0)=\Phi_0/[2\pi\xi^2(0)]$. 

In summary, we found that centrosymmetric La$_3$Pt$_4$ exhibits superconductivity below 0.51~K, 
while trying to synthesize LaPdC$_2$ and LaPtC$_2$ using arc melting. 

\begin{acknowledgment}
We thank Y. Yamaoka, T. Iye, S. Kitagawa, K. Ishida, and H. Sawa for fruitful discussion. 
This work is supported by a Grant-in-Aid from the Global COE program 
``The Next Generation of Physics, Spun from Universality and Emergence'' 
from the Ministry of Education, Culture, Sports, Science, and Technology (MEXT) of Japan, 
and by the ``Topological Quantum Phenomena'' Grant-in Aid for Scientific Research on innovative 
Areas from MEXT of Japan. G.E. is also supported by JSPS.
\end{acknowledgment}


\end{document}